\documentclass[%
 reprint,
 aps,
 prb,
 epsf,
]{revtex4-2}

\usepackage{graphicx}
\usepackage{dcolumn}
\usepackage{bm}
\usepackage{hyperref}
\usepackage{mathtools}
\usepackage{xcolor}
\usepackage{ulem} 
\usepackage{float}
\usepackage{braket}

\begin{document}

\title{Noise and fluctuations in nanoscale gas flow}

\author{J. Dastoor}
\author{D. M. Willerton}
\author{W. Reisner}
\author{G. Gervais}

\affiliation{%
 Department of Physics, McGill University, Montreal, Quebec H3A 2T8, Canada
}%

\date{\today}

\begin{abstract}
We theoretically calculate the fundamental noise that is present in gaseous (dilute fluid) flow in channels in the classical and degenerate quantum regime, where the Fermi-Dirac and Bose-Einstein distribution must be considered. Results for both regimes are analogous to their electrical counterparts. The quantum noise is calculated for a two terminal system and is a complicated function of the thermal and shot noise with the thermal noise dominating when $2k_BT\rho \gg m\Delta P$ and vice versa. The cumulant generating function for mass flow, which generates all the higher order statistics related to our mass flow distribution, is also derived and is used to find an expression for the third cumulant of flow across a fluidic channel.
\end{abstract}
\maketitle

\section{\label{sec:intro}Introduction}

     Nanoscale fluid transport in dilute (gaseous) regimes is of broad fundamental and engineering importance, relevant in diverse scenarios ranging from understanding fluid flow in quantum and classical regimes to industrial applications involving gas processing \cite{bao}.  From a fundamental point of view, improvements in nanofabrication have enabled the production of nanochannels and nanopores with well-defined nanometric dimensions that can be used to verify classical theories for gas flow in free-molecular transport (Knudsen) regimes \cite{huber,savard,velasco}; these measurements have been extended cryogenically to explore transport of quantum fluid phases of $^4$He \cite{gervaispore4He} where it is expected that a one-dimensional many-body quantum state would form \cite{gervaispore4He2}.   From an engineering point of view, nanochannels and nanoporous materials, due to their high surface to volume ratio and pore sizes below the molecular mean free path and/or approaching molecular dimensions \cite{wang}, have excellent absorptive properties and can exhibit size-based molecular sieving \cite{wang,Geim, Ding}, useful for applications in gas separation \cite{Ding} and catalysis \cite{Li}.\\
     
    Most experimental and theoretical efforts devoted to characterizing nanoscale gas transport have focused on modeling the gas mass flow-rate $Q$,  {\it e.g.} \cite{scorrano,shen}.  In analogy to the case of electrical transport, this is given by $Q = G\Delta P$, where $Q$ is the mass flow, $G$ is the flow conductance, and $\Delta P$ is the pressure difference across the channel or pore \cite{savard}.   However, the gas mass flow-rate is not the only quantity of interest that can be extracted via monitoring a given mass flow channel.  Just as is the case for electrical current, statistical fluctuations in the mass flow will exist (mass flow noise).  These fluctuations are also of fundamental interest, for example providing new information about a system's fluidic properties in both classical and quantum regimes -- {\it e.g.}, its transmission properties or flow limitations.  The mass flow noise will also have practical implications, limiting applications where mass-flow rate is used as a sensor by creating a noise floor that observable signals need to exceed.   In addition, there might be scenarios where the degree of mass flow noise present could itself constitute the signal of interest.   Finally, gas-flow fluctuations might affect the performance of gas based separation or catalysis devices, for example statistical fluctuations in the stream of a low concentration catalysis or inhibitor species might lead to large fluctuations in output.\\
    
Considerable effort has been devoted in the past to improve our understanding of electrical noise; much of this insight can be adopted to the closely analogous case of mass flow noise.  White current fluctuations arise due to thermal energy (Johnson-Nyquist noise) and the discrete nature of electrical charge (shot noise) \cite{wavepacket}. These two noise sources are classically distinct, yet become interlinked in the quantum regime. The development of quantum shot noise theory has seen new applications in distinguishing particles from waves and future proposals for new entanglement detectors \cite{entanglement}, and has led to the spectacular experimental validation of the effective quasiparticle charges of electrons confined to two dimensions in the fractional quantum Hall regime \cite{quasiparticle,heiblum}. Somewhat surprisingly these sources of noise that set fundamental limits in terms of signals in the case of dilute mass flow has been neglected, and so here we propose a theoretical calculation for the thermal noise of an ensemble of particles, forming a dilute gas, by adapting well-defined techniques developed for electrical noise. We also calculate the noise associated with a directed flow (shot noise) in the classical regime and then adapt our discussion to include quantum effects by combining thermal and shot noise. This quantum noise is sometimes referred to as the quantum shot noise and will be derived by adapting Martin and Landauer's wave-packet approach for electrical noise to mass flow noise \cite{wavepacket}. This approach is chosen because it allows us to develop a quantum noise expression based on a similar process to our derivation of classical noise while straightforwardly incorporating the relevant quantum mechanical considerations.  All our expressions are derived based on the assumption that the noise is distributed equally over all frequencies (i.e. that it is \textit{white}) and conforms with a Gaussian distribution, which is expected based on analogy with the electrical noises.\\
    
    We produce general results from the fundamental flow equation, $Q = G\Delta P$, where $Q$ is the mass flow, $G$ is the flow conductance, and $\Delta P$ is the pressure difference across the channel. In doing so, we neglect any consideration of turbulence since our concern is in small dilute fluidic systems with very low Reynolds numbers, $Re  \ll  2000$. It should however be noted that under certain conditions even small mesoscopic systems may be subject to turbulent flow, e.g. \cite{microchannels, microchannels 2, micropipes}, and in such cases a new approach may  be needed.  We have also confined our analysis to idealized fluidic channels in which electromagnetic effects are negligible. We note that in many nanofluidic systems, charges on mass carriers and surface effects can heavily influence the mass flow and a new approach may be needed to address these cases as well \cite{gogoi,kavokine}. The noise, $\delta Q$, will be calculated as a mean squared fluctuation, $\langle \delta Q^2\rangle$, where $\langle \dots\rangle$ refers to an average with respect to time. Recent work found that $G$ is quantized in units of $2m^2/h$, where $m$ is the mass of a fluid particle and $h$ is Planck's constant \cite{Lambert2008, quantized}. The resemblance with the quantum of electrical conductance ($2e^2/h$ where $e$ is the electrical charge) highlights the close analogy between mass flow and charge flow. \\
    
    Other sources of noise in mesoscopic fluidic systems, such as the irregular motion of impurities, also become important to consider when building sensitive devices. While these sources are not explored here, the cumulant generating function of the quantum white noise is calculated to allow for easy combination with other independent noise sources in future scenarios. In addition, we also derive an expression for the third cumulant of quantum white noise. With growing interest in the theory of full counting statistics, these will both be valuable tools for future work. Finally, we verify our results using the fluctuation dissipation theorem, which describes the relation between random fluctuations of a system at equilibrium to a small perturbation. The theorem states that fluctuations occurring in equilibrium, {\it i.e.} in the absence of a net mass flow, are proportional to the channel conductance \cite{fractional}. This provides a baseline test for our results.

\section{\label{sec:classnoise}Classical Noise}

Noise in the classical regime is separated into thermal noise and shot noise contributions. In dilute fluidic systems, the shot noise originates from the discrete nature of an average mass flow signal and is thus formally only present out of equilibrium. The thermal noise, however, is expected to exist even at thermodynamic equilibrium as it is due to the innate, random motion of particles that is implied by the kinetic theory of gases and the Maxwell-Boltzmann distribution. We may conceptualize the ensuing random passage of particles across the pore as comprising an instantaneous flow rate that averages to zero over long times, i.e. $\braket{Q} = 0$, and is thus permitted to occur even in the absence of a net pressure differential, $\Delta P$. \\

We begin by considering the thermal fluctuations in a system at equilibrium by noting that when both terminals are in equilibrium with each other, there is no net energy transfer within the system and we can use the analogy of a standing wave in an open pipe. We know from equipartition that each standing wave mode has two degrees of freedom and therefore each mode has an average energy of $k_BT$, where $k_B$ is Boltzmann's constant and $T$ is temperature.\\
    
    \vspace{0mm}
    \begin{figure}[!ht]
        \centering
         \includegraphics[width=0.83\columnwidth]{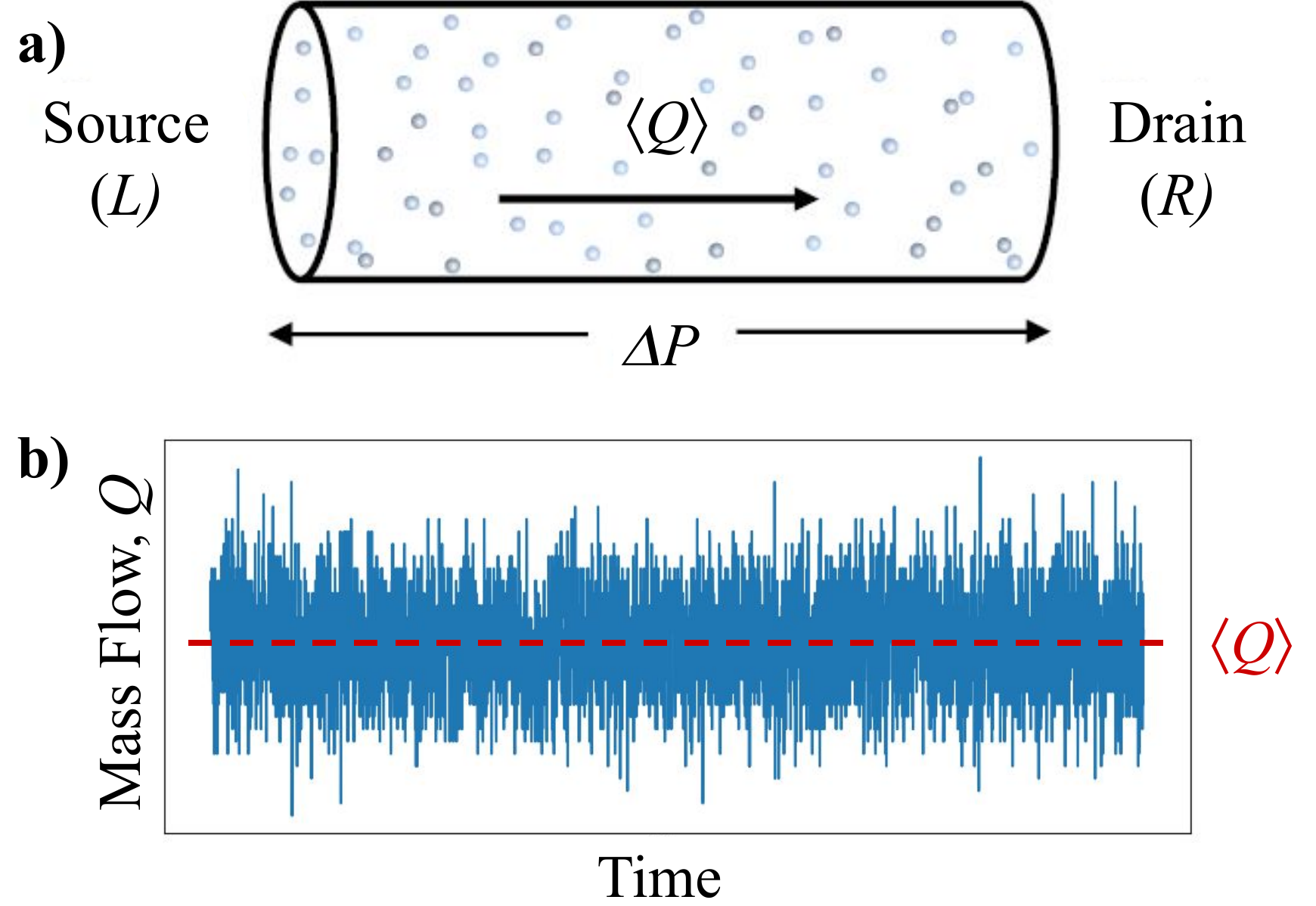}
        \vspace{-1mm}
        \caption{\textbf{a)} A cartoon of the flow through a cylindrical fluid channel. Flow is from the left side to the right side (source to the drain).\label{fig:flow} \textbf{b)} Cartoon example of mass flow fluctuations about an average flow signal, $\langle{Q}\rangle$, occurring in a dilute nanochannel.}
    \end{figure}
    \vspace{0mm}

   The total average energy of our system is found as $\Delta j k_BT$, where $\Delta j$ is the range of modes that our system can occupy. The occupied range of modes is characterised by the time taken for a particle to pass through the channel and by the frequency range, $\Delta \nu$, that the standing waves can occupy. An equation for the average power can now be written using our previous results and the frequency of oscillation of our standing wave. Lastly, we equate this power expression with the instantaneous thermodynamic power of the fluid and derive an expression for the thermal noise in a fluid channel:
    \begin{equation}\label{eq:class_thermal}
        \langle \delta Q^2 \rangle = 4k_BTG\rho \Delta \nu,
    \end{equation}
    where $\rho$ is the mass density of the fluid. The fluctuation dissipation theorem is satisfied since the noise is proportional to the conductance. Note that equation \ref{eq:class_thermal} shows that thermal noise exists even when $\langle Q \rangle = 0$ (at equilibrium). The Johnson-Nyquist expression for the electrical current thermal noise, $4k_BTR^{-1}\Delta\nu$ (where $R$ is the electrical resistance), is analogous to equation \ref{eq:class_thermal} \cite{nyquist}.\\

In the case that there exists a nonzero pressure differential across the pore, there will be a net
flow of mass from one reservoir to the other and equation \ref{eq:class_thermal} no longer suffices
to describe the fluctuations. To formulate a noise expression that accommodates an
arbitrary pressure differential across the pore, it is necessary to model the mass flow as consisting in
a stream of discrete particles, much like an electrical current. We further note that, in the free-molecular flow regime, interparticle collisions are negligible and we can model the flow of particles as a Poisson process. We may then loosely follow van der Ziel’s derivation for electrical shot noise, adapting the approach where necessary,
to establish a comprehensive expression for the mass-flow fluctuations \cite{vanderziel}.\\

Suppose that on either side of a nanopore we hold reservoirs at fixed pressures, $P_A$ and $P_B$, such that 
\begin{equation}
    P_A =  \Delta P+ P_0 ;\; P_B = P_0,
\end{equation}
where we have identified reservoir $A$ as the region of higher pressure and have used the pressure of $B$ to mark a baseline, $P_0$. Particles are transmitted across the pore in both directions, each with some associated instantaneous rate of occurrence, $r(t)$, which fluctuates in time. Defining $N$ to be the net number of particles passing through the pore in the direction of $A$ to $B$, we have
\begin{equation}
    N = \int_0^\tau [r_{A\rightarrow B}(t)  -  r_{B \rightarrow A}(t)]dt,
\end{equation}
for some time interval, $\tau$. Note that a negative $N$ signifies a net passage of particles from $B$ to $A$. We may also define the fluctuation in $N$ as the instantaneous deviation from its average, 
\begin{equation}
    \delta N = N - \braket{N},
\end{equation}
with 
\begin{equation}
    \braket{N} = \braket{r_{A\rightarrow B}}\tau  - \braket{r_{B\rightarrow A}}\tau.
\end{equation}
The averages in the above expression may be interpreted as either ensemble or time averages, since these are equivalent for an Ergodic process. We now define an additional random variable, $\delta R_\tau$, corresponding to fluctuations in the net rate of particle transmissions as
\begin{equation}
    \delta R_\tau = \frac{\delta N}{\tau}.
\end{equation}
Noting that the variance of $N$ is defined as $Var(N) = \braket{\delta N^2}$, we may write that 
\begin{eqnarray}
\label{eq:7}
    \braket{\delta R_\tau^2} &&= \frac{Var (N)}{\tau^2} = \frac{Var (N_{A\rightarrow B} - N_{B \rightarrow A})}{\tau^2}\\
    &&= \frac{ Var(N_{A\rightarrow B}) + Var(N_{B\rightarrow A})}{\tau^2},
\end{eqnarray}
where the final step follows from the fact that the variance for a difference on two Poisson variables is just the (positive) sum of the variances associated with each individual process. Additionally, because the variance of a Poisson process is equal to its mean, we have

\begin{equation}
   \braket{\delta R_\tau^2}   = \frac{\braket{r_{A\rightarrow B}} + \braket{r_{B\rightarrow A}}}{\tau}.
\end{equation}
Applying the Wiener-Khintchine theorem allows us to extract the zero-frequency component of the noise spectral density as
\begin{equation}
    S_R(0) = \lim_{\tau\rightarrow \infty} 2\tau \braket{\delta R_\tau^2} = 2\braket{r_{A\rightarrow B}} + 2\braket{r_{B\rightarrow A}},
\end{equation}
 which, for white noise, suffices to describe the entire spectrum. We now make the conversion to units of mass-flow fluctuations by multiplying the spectral density by the mass of the fluid particles squared:
 \begin{equation}
     S_Q(0) = 2m\braket{Q_{A\rightarrow B}} + 2m\braket{Q_{B\rightarrow A}},
 \end{equation}
 where we have distributed one factor of $m$ into each average to pass from particle flow rates to mass flow rates. We may alternatively write the above expression as
 \begin{equation}\label{eq:density}
     S_Q(0) = 2mGP_A + 2mGP_B = 2mG\Delta P + 4mGP_0,
 \end{equation}
which is legitimate because the transmission events are statistically independent and we may consider $P_A$ and $P_B$ separately, each constituting an effective pressure differential across the pore. The first term on the right contains the factor $G\Delta P $, which we know to be the average net mass flow, $\braket{Q}$. If we make this substitution and also replace $P_0$ with $nk_BT$ via the ideal gas law, then we may multiply through by an arbitrary frequency bandwidth, $\Delta v$, to arrive at the result 
\begin{equation}
    \braket{\delta Q^2} = 2m\braket{Q}\Delta v + 4Gk_BT\rho_0\Delta v,
\end{equation}
with $\rho_0$ being defined as a baseline mass density that exists across both reservoirs. When cast in this form, the above expression lends itself to a straightforward analogy with electrical circuits. Firstly, we note that when $\braket{Q} = 0$, or equivalently when the two reservoirs exist in thermodynamic equilibrium, we recover equation \ref{eq:class_thermal}, which was previously likened to the Johnson-Nyquist thermal noise in electrical resistors. We also note that the first term on the right may be identified with Schottky's result for electrical shot noise, $\braket{\delta I^2} = 2e\braket{I}\Delta v$ \cite{schottky}. \\

\begin{figure}[!ht]
        \centering
         \includegraphics[width=1\columnwidth]{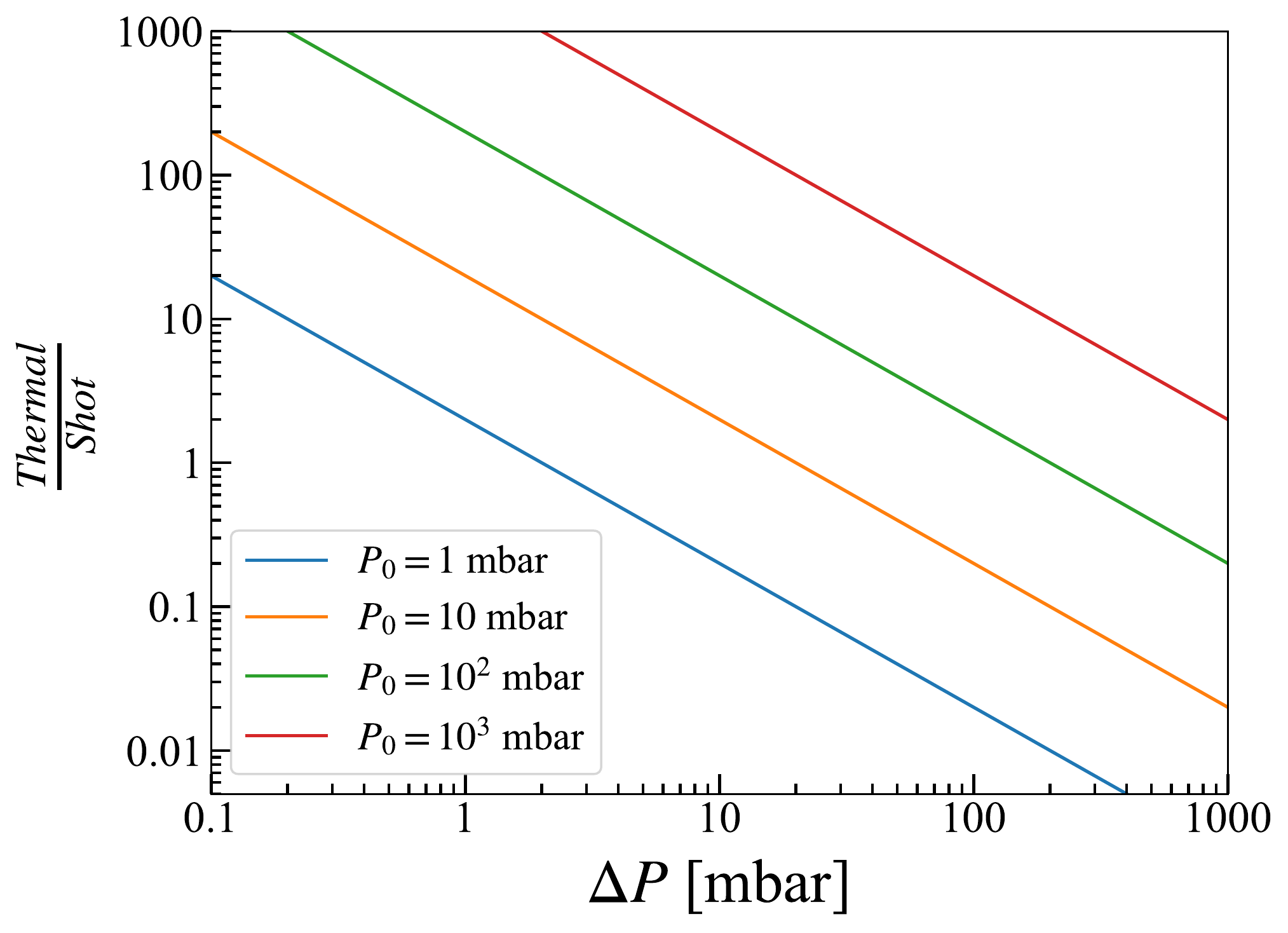}
        \vspace{-8mm}
        \caption{ The ratio of thermal noise to shot noise plotted against the pressure differential, $\Delta P$, across a cylindrical nanopore. Various baseline pressures, $P_0$, have been distinguished by line colour. The displayed range of pressures have been somewhat arbitrarily chosen based on typical conditions for free-molecular $^4$He gas flow in nanopores of $\leq$ 100 nm diameters, such as those used in \cite{savard}. We also reiterate that, strictly speaking, this result is only valid in the free-molecular flow regime.\label{fig:jonson_shot}}
    \end{figure}

If we interpret the first term in equation \ref{eq:density} as the \textit{mass-flow} shot noise and the second term as the \textit{mass-flow} thermal noise, then we can construct the unitless ratio 
\begin{equation}
    \frac{\braket{\delta Q^2_{therm}}}{\braket{\delta Q^2_{shot}}} = \frac{2P_0}{\Delta P},
\end{equation}
to identify the conditions under which each noise source is expected to dominate. From Figure \ref{fig:jonson_shot} it is clear that when $\Delta P$ is sufficiently small compared with $P_0$, the thermal fluctuations are dominant. This is expected because the net flow, which gives rise to the shot noise, will be small compared with the opposing flows resulting from $P_0$ and thus contribute much less to the overall fluctuations. When $\Delta P$ is large compared with $P_0$, the reverse is true and the shot noise is expected to dominate.

\section{\label{sec:quantnoise}Quantum Noise}
    In this section we consider a system in which quantum effects are taken into account and the particles may no longer be treated independently. In this case,  the thermal noise and shot noise are interlinked, and there is an inherent probability associated with transmission through the channel. We theoretically calculate the noise for a two-terminal cylindrical fluid channel although this approach can easily be generalised to a multi-channel system. The shot noise can easily be adapted at zero temperature to account for the transmission probability, $D$, by recognising that the Poissonian noise distribution becomes binomial. We also assume $D$ to be energy independent for the rest of this paper. In this case, the fluctuations are given by:
    \begin{equation}\label{eq:quant_shotT=0}
        \langle \delta Q^2 \rangle = 2m\langle Q\rangle (1-D) \Delta \nu,
    \end{equation}
    where $Q$ is now defined as the outgoing flow from the channel and hence, implicitly absorbs a transmission probability factor. Note that the ingoing and outgoing flow now differ by a factor $D$.\\
    
\begin{figure}[!ht]
        \centering
         \includegraphics[width=1\columnwidth]{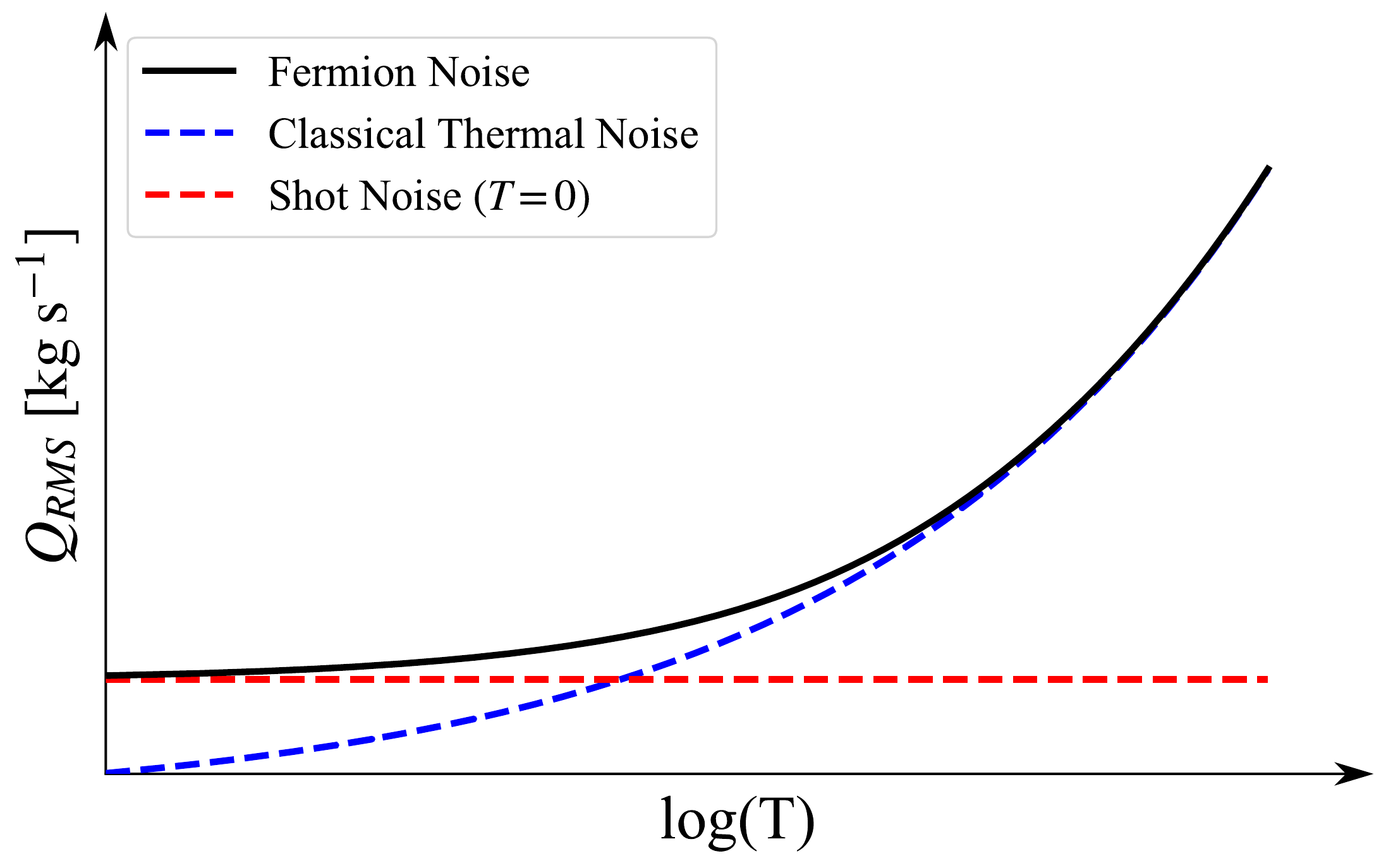}
        \vspace{-3mm}
        \caption{The root mean square (RMS) mass flow fluctuations for fermions in a two terminal system plotted against the log of temperature using arbitrary parameters. The classical thermal noise and zero-temperature shot noise have been plotted as dashed lines for comparison. Note that the intercepts with the $Q_{RMS}$ axis occur at zero and the square root of equation \ref{eq:quant_shotT=0}.  \label{fig:fermionfinalnoise}}
    \end{figure}

    When $T>0$, we must consider thermal fluctuations in the incoming and outgoing flow. Assuming thermal equilibrium, the occupation of states at the source and drain are governed by the Fermi-Dirac distribution for fermions, and the Bose-Einstein distribution for bosons. Both will be referred to as $f$ in their respective contexts, with chemical potentials $\mu_L$ and $\mu_R$, referring to the left and right side of the channel, where $(\mu_L-\mu_R) = m\Delta P/(\rho)$. Transport from the left to right is defined as positive in quantized units of $G$ where \cite{Lambert2008}:
    \begin{equation}\label{eq:quant_G}
        G = \frac{Q}{(\mu_L-\mu_R)\rho/m} = \frac{2m^2}{h\rho}D.
    \end{equation}
    We now adapt Landauer's wave packet approach for electric circuits to mass flow to find an expression for the noise. In this approach, transmission and reflection are characterised by wave packets that are emitted from the source and the drain at a constant rate and each contain one quantum mechanical state \cite{wavepacket}. Note that due to the Pauli exclusion principle, only two fermions can occupy this state (opposite spins) whereas there is no restriction for bosons. Packets are assumed to be emitted in phase and simultaneously from the source and drain, such that a transmitted wave from the source is able to map onto the same state as a reflected wave from the drain and vice versa. This method allows us to consider particles moving against the pressure gradient (which  is more probable at low pressures). Using counting statistics we find:
    \begin{eqnarray}\label{eq:quant_noiseIntegral}\nonumber
        \langle \delta Q^2\rangle = \frac{4m^2}{h}\Delta \nu\int_0^\infty dE &&\{D[f_L(1\mp f_L) + f_R(1\mp f_R)] \\
        &&\pm D(1-D)(f_L-f_R)^2\},
    \end{eqnarray}
     where $f_L$ and $f_R$ denote the distribution of particles at the source and drain respectively. Note that the upper sign is for fermions and the bottom is for bosons. The integral serves to include wave packets of all energies. For bosons, the integral diverges, as is also seen for bosons in electric circuits \cite{fractional}. Hence, for the remainder of this section we will focus on fermions. For fermions, we have the exact result:
    \begin{eqnarray}\label{eq:quant_fermions}\nonumber
        \langle \delta Q^2\rangle= &&4k_BTG\rho \Delta\nu D\\
        &&\;\;+ 2m\langle Q\rangle (1-D)\Delta\nu \coth{\left(\frac{m\Delta P}{2kT\rho}\right)}.
    \end{eqnarray}
    The first term is our classical thermal noise with a transmission factor, whereas the second term is our classical shot noise with a complicated cutoff factor. When $k_BT\rho \gg m\Delta P$, the hyperbolic cotangent is approximated by the inverse of its argument and we recover our classical thermal noise expression given by equation \ref{eq:class_thermal}. Hence, the quantum noise satisfies the fluctuation dissipation theorem as the equilibrium noise is proportional to the conductance. When $k_BT\rho \ll m\Delta P$, the hyperbolic cotangent is approximately one and the second term dominates, so we recover our zero-temperature shot noise expression given by equation \ref{eq:quant_shotT=0}. This is further shown in Figure \ref{fig:fermionfinalnoise}.
    
    \vspace{0mm}
    
\section{\label{sec:cumulants}Cumulants}
    The cumulants provide an alternative description of a random variable to the probability distribution, where the first cumulant is the mean, the second is the variance, the third describes the asymmetry or skewness of the distribution, etc. For statistically independent variables, the cumulants are additive, making them extremely useful when seeking to sum over multiple independent sources of noise \cite{FullCounting}. For this reason, we now provide a reformulation of our results for the quantum noise in terms of a cumulant generating function. \\
    
    The $n$-th cumulant, $\kappa_n$, can be found by evaluating the $n$-th derivative of the cumulant generating function (CGF), $K(t)$, at $t=0$ \cite{FullCounting}. For fermions, the CGF takes the form:
    \begin{eqnarray}\label{eq:cum_K(t)}\nonumber
        K(t) =&&\frac{4}{h\Delta\nu} \int_0^\infty dE\ln\{\left(e^{m\Delta\nu t}-1\right) f_L D(1-f_R) \\
        &&\;\;\;\;\;\;\;\;+\left(e^{-m\Delta\nu t} - 1 \right) f_RD(1-f_L)+1\}.
    \end{eqnarray}
    The first two factors represent transport from the left to right or right to left of the channel, respectively. Note that if we attempt to derive a similar expression for bosons, all our integrals diverge and are hence, meaningless. Focusing on fermions, the exponential mass factor is explained by recognising that the $n$-th cumulant can be thought of as a pairing of $n$ independent variables, where each variable contributes a mass factor. The frequency factor results from the Fourier transform. Under the zero temperature limit, equation \ref{eq:cum_K(t)} becomes:
    \begin{equation}\label{eq:cum_KT=0}
        \lim_{T \to 0} K(t) = \frac{4m}{\rho h\Delta\nu}\Delta P \ln \left\{1-D+De^{m\Delta\nu t} \right\}.
    \end{equation}
    This expression is, as expected, reminiscent of the well-defined CGF of the binomial distribution, $n\ln\{1-p+pe^t\}$, where $n$ is the number of trials and $p$ is the probability \cite{FullCounting}. We are also able to use equation \ref{eq:cum_KT=0} to recover our zero-temperature shot noise, given by equation \ref{eq:quant_shotT=0}, from $K''(0)$. On the other hand, if we calculate $K''(0)$ when $T>0$, we recover the fermion variance given by equation \ref{eq:quant_fermions}. \\
    
    Utilizing our fermion CGF, we can further improve our understanding of the distribution of mass flow by calculating the skewness (or third cumulant).
    \begin{eqnarray}\nonumber
        &&\kappa_3 = \frac{1}{2}mGk_BT(\Delta \nu)^2 (1-D)\mathrm{csch}^2\left( \frac{m\Delta P}{2k_BT\rho}\right) \\ 
        &&\nonumber\times\left\{ 6D\rho \sinh{\left(\frac{m\Delta P}{k_BT\rho}\right)} +(2D-1)\frac{m\Delta P}{k_BT}\cosh{\left(\frac{m\Delta P}{k_BT\rho}\right)}\right. \\ 
        &&\hspace{0.5\columnwidth}\left. - (1+4D)\frac{m\Delta P}{k_BT}\right\}.
    \end{eqnarray}
    The expression above is complex, and no longer similar to our usual thermal and shot noise. Under the limit $k_BT\rho\gg m\Delta P$, our expression for $\kappa_3$ becomes $4(G/\Delta P)(\rho k_BT)^2 (2-D)(1-D)(\Delta\nu)^2$ and in the opposite limit it becomes $m^2\langle Q\rangle (1-D)(2D-1)(\Delta\nu)^2$. Under the classical limit, approximately when $D=1$, we see that $\kappa_3=0$. Hence, departure from the classical regime can be detected as $\kappa_3\neq0$. This further shows the non-Gaussian like nature of mass flow noise in the quantum regime, even under the assumption of zero frequency.
    
\section{\label{sec:discussion}Discussion and Outlook}
    Our results demonstrate that flow in nanofluidic channels is subject to innate mass-flow fluctuations, which vary depending on the conditions of the system and may be a relevant consideration in the study and applications of nanofluidics. In the previous sections we have derived classical and quantum expressions for the white noise prevalent in mass flow in two-terminal fluid channels. We have also determined the CGF for fermions and provided expressions for $\kappa_3$. Notably, even though we have only assumed white noise, our expressions should still be valid at low frequencies where $k_BT\gg h\nu$. However, at high frequencies where $k_BT\approx h\nu$, one would expect quantum effects and must replace the classical expression for average energy $k_BT$ with its quantum version: $h\nu /[\exp{(h\nu/k_BT)}-1]$. Similar substitutions have been made for the electrical case of a quantum point contact, and have been found to agree with experiments \cite{QPC}.\\
    
    Generalizations to multi-channel systems is possible by summing over the different transmission probabilities corresponding to each channel. The number of channels in a typical pipe is $A/{\lambda_F}^2$, where $A$ is the cross-section of the pipe and $\lambda_F$ is the Fermi wavelength. Hence, with larger channels this will become more important. Furthermore, there are a number of different sources of noise and types of noise that have yet to be studied for mass flow. For instance, there may be extrinsic sources of noise that are sensitive to boundary layer effects and system imperfections. These can be studied with a specific system in mind. Additionally, noise inversely proportional to frequency, traditionally called $1/f$ noise, is common in electric circuitry and can also be studied here. Turbulent flow might also impact our expressions for noise and should be studied in more detail.\\
    
        Note that our calculations assume that the particles colliding with device surfaces undergo specular rather than diffuse reflections.  This assumption might not hold for measurements performed in long nanochannels fabricated via classic nanomachining approaches as the device surfaces in these cases are not atomically smooth \cite{Qian}.  However, there is extensive interest in gas transport in materials possessing atomically smooth surfaces \cite{wang}, such as channels formed from carbon nanotubes \cite{bakajin}, graphene \cite{Geim}, MoS$_2$ and h-BN \cite{keerthi}. These materials lead to enhanced gas transport while not loosing their separation selectivity \cite{wang}.\\
    
    It is also interesting to consider real world applications where both ends of the channel might not be at thermal equilibrium. A temperature difference can be incorporated into our method by adapting $f_L$ and $f_R$. However, adding a more complex gradient is an interesting problem that requires more theoretical consideration. In this case, cumulants could be used to sum up many distributions and account for varying temperature differences.
    
\section{\label{sec:conclusion}Conclusion}
    We have theoretically calculated the classical and quantum noise, and quantum  CGF, for mass flow in a dilute fluid channel. The result is found to be analogous with previous calculations for electrical noise and obeys the fluctuation dissipation theorem. We have also used the CGF to determine the third cumulant which, even at zero frequency, is non-zero. This shows the non-Gaussian nature of mass-flow noise.\\
    
    Although our results are mathematically similar to the electric case, the adaptation of methods for electrical noise to mass flow is important.  The classical, free-molecular flow noise can be thought of as the sum of the full shot noises associated with two opposing mass flow currents, which each originate from the underlying Poisson statistics of the discrete mass carriers. The quantum noise may be similarly understood, with the acknowledgement that in this case the statistics of the particles must conform with the corresponding quantum mechanical prescription, thus leading to interparticle interactions which were not accounted for in the classical case.  The theoretical prescription presented here can be used to calculate the theoretically minimum noise expected in dilute fluidic channels, or help justify experimentally observed signal fluctuations.\\
    
     We believe that these insights are important to understand smaller and more complex fluid systems and thus may be of use in areas of nanofluidics where noise considerations are of crucial importance, such as in certain sensing technologies \cite{nanopore noise}. We hope that this work may serve as a stepping stone for more elaborate noise sources to be considered in the future. For instance, future work may include an extension of our classical expressions into other gas flow regimes, such as into the transition and continuum regimes, since strictly speaking we have confined our classical analysis to free-molecular flow. The noise theory of dilute fluid channels should also be extended to high frequencies, considering other sources of noise, and adapting the results to fit more realistic systems with temperature gradients and numerous channels.

\section*{Acknowledgments}
   This work has been financially supported by NSERC (Canada), the New Frontier in Research Fund (Canada), FRQNT (Qu\'ebec) and the McGill Tomlinson fund.

\end{document}